\documentclass[aps,prx,twocolumn, groupedaddress]{revtex4-2}
\usepackage{graphicx}
\usepackage{xcolor, amsmath, comment}

\begin{document}

\title{Swarming and Flocking Unified Through Aggregation}

\author{Joao Lizárraga}

\author{Marcus de Aguiar}
 \email{aguiar@ifi.unicamp.br}
\affiliation{
	Instituto de Física Gleb Wataghin, Universidade Estadual de Campinas, Unicamp 13083-970, Campinas, São Paulo, Brazil
}

\date{\today}

\begin{abstract}
Natural flocks (aligned) and swarms (non-aligned) both exhibit features of near-criticality, challenging their treatment as two ends of the same phase transition. We present a model for the aggregation of active individuals, in which their velocities align as a byproduct of achieving stable cohesion. In our framework, individuals move in open space and possess differing self-propelling velocities. Furthermore, velocity fluctuations are triggered by individual errors when following the aggregation rules. Notably, the system exhibits scale invariance, which is shown to be rooted in the model's definition---a feature that we label as \textit{structural criticality}. Finally, we show the emergence of a striking regime where spatial and orientational coherence decouple. That is, the system can achieve states of \textit{high} and \textit{low} polarization while maintaining spatial homogeneity.
\end{abstract}
\maketitle

\section{Introduction}
Swarming and flocking are ubiquitous. Although commonly associated with insects~\cite{yates2009inherent, sayin2025behavioral} and birds~\cite{ballerini2008interaction, bialek2012statistical}, similar behaviors are also exhibited by fish~\cite{zada2024development, puy2024selective} and mammals~\cite{gomez2022intermittent, yan2024collective}, among others. In these systems, groups form mediated by the self-organization of individuals who seem to follow similar interaction rules. One of these, the alignment of orientations, paved the way for the theoretical study of collective phenomena, as it prompted the formulation of the seminal Vicsek model~\cite{vicsek1995novel}. 

In Vicsek's framework, individuals with identical velocities are set to move in a periodic space; coherence is promoted by mirroring neighboring orientations. Then, depending on the noise level and individual density, a (pseudo) second order transition arises between aligned and non-aligned states ~\cite{chate2008collective, baglietto2008finite}. Its success in reproducing these behaviors led to its subsequent adoption as an almost universal tool for addressing related problems~\cite{peruani2008mean, toner1998flocks, farrell2012pattern, miguel2018effects, gregoire2004onset, barberis2016large, mugica2022scale, cavagna2023natural, shi2025collective}. However, once discussions extend beyond the study of \textit{prototypical} systems, its applicability gets challenged~\cite{sayin2025behavioral}.

The aim of this paper is to introduce an alternative mechanism to that of velocity alignment: aggregation. Although we cannot state with certainty which of these phenomena is more important, it is undeniable that aggregation is common to a vast number of collective systems. To search for food~\cite{schrandt2015facilitation}, to look for mating partners~\cite{attanasi2014collective}, to face predators~\cite{cavagna2014bird}, and even during the early development of living systems~\cite{tan2022odd}; once grouped, individuals behave as a whole. Although our model differs fundamentally from Vicsek's, velocity alignment can still be achieved; even without the explicit definition of interactions for that purpose. In a similar vein, we introduce the idea of \textit{behavioral imperfections}, which represent individual errors when following the aggregation rules rather than \textit{sensing} inaccuracy. 

This paper consists of five sections. The aggregation mechanism, as well as implications regarding the noise and coupling strengths, is described in Sec.~\ref{sec:sec02}. The system's collective states, contrasting spatial and orientational coherence, are presented in Sec.~\ref{sec:sec03}. In Sec.~\ref{sec:sec04}, we present numerical and analytical arguments on the system's \textit{structural criticality}. Finally, details on the system's limitations and outlook are discussed in Sec.~\ref{sec:sec05}.

\section{The aggregation mechanism}
\label{sec:sec02}
In a system of $N$ particles, the dynamics of the $i$-th individual are governed by
\begin{equation}
\dot{\vec{r}}_i = \vec{f}_i - {\alpha}_i \nabla_{\vec{r}_i} V_i(\vec{r}),
\label{eq:pot}
\end{equation} 
where $\vec{r}_i$ denotes its $n$-dimensional position. The influence exerted by its own state and those of its neighbors is characterized by $V_i(\vec{r})$, which is scaled by the coupling strength $\alpha_i$. The argument $\vec{r}= (\vec{r}_1, \vec{r}_2,\dots ,\vec{r}_N)$ indicates that $V_i$ depends on the coordinates of all individuals. The term $\vec{f}_i$, considered to be constant for simplicity, represents the individual's velocity when isolated (i.e., its self-propulsion).

The aggregation phenomenon we study is driven by the potential 
\begin{equation*}
V_i(\vec{r}) = \frac{r_i^2}{2} - \frac{1}{N}\sum_{j= 1}^N\vec{r}_i\cdot \vec{r}_j,
\end{equation*}
where $r_i= |\vec{r}_i|$. Depending on the coupling strength ($\alpha_i$), then, this parabolic well drives $\vec{r}_i$ either to instability or to a specific \textit{bounded} state. Moreover, when plugged into Eq.~\eqref{eq:pot}, it leads to
\begin{equation}
    \dot{\vec{r}}_i = \vec{f}_i + \alpha_i\left(\langle\vec{r}\rangle - \vec{r}_i\right),
    \label{eq:mf}
\end{equation}
with $\langle\vec{r}\rangle$ being the population's center of mass (mean position).

\subsection{Coupling heterogeneity}
Despite the simplicity of Eq.~\eqref{eq:mf}, the system's complexity lies in its governing foundations: coupling strengths can be different ($\alpha_i\neq\alpha_j$). Taking this into account, and averaging both sides of Eq.~\eqref{eq:mf}, we obtain
\begin{equation}
    \langle\dot{\vec{r}}\rangle = \langle\vec{f}\rangle - \mathrm{Cov}(\alpha, \vec{r}),
    \label{eq:extavg}
\end{equation}
where $\mathrm{Cov}(\alpha, \vec{r}) = \langle \alpha\vec{r}\rangle - \langle\alpha\rangle\langle\vec{r}\rangle$. This early derivation allows us to make a first prediction on the system's long-term behavior. Specifically, that the instantaneous coherence between individual velocities is characterized by
\begin{equation}
|\langle\dot{\vec{r}}\rangle|^2 = |\langle\vec{f}\rangle - \mathrm{Cov}(\alpha, \vec{r})|^2. 
\label{eq:op01}
\end{equation}
That is, as long as the system maintains statistical homogeneity, a state where the collective order is zero can only be achieved if $\langle\vec{f}\rangle = \mathrm{Cov}(\alpha, \vec{r})$.

Using the system's mean velocity [Eq.~\eqref{eq:extavg}] once more, we can rewrite Eq.~\eqref{eq:mf} as
\begin{equation}
    \dot{\vec{u}}_i = \vec{f}_i - \langle\vec{f}\rangle -\alpha_i\vec{u}_i+\mathrm{Cov}(\alpha, \vec{r}),
    \label{eq:flucgen}
\end{equation}
which characterizes the evolution of individual fluctuations ($\vec{u}_i = \vec{r}_i - \langle\vec{r}\rangle$). On this basis, it is trivial to notice that, if the system reaches its equilibrium ($\dot{\vec{u}}_i = 0$), the distance of each individual towards the center of mass will be given by
\begin{equation}
    \vec{u}_i = \frac{1}{\alpha_i}\left[\vec{f}_i - \langle\vec{f}\rangle + \mathrm{Cov} (\alpha, \vec{r})\right].
    \label{eq:conv}
\end{equation}
Once settled in this state, individual velocities are \textit{identical} to that of the center of mass: $\dot{\vec{r}}_i = \langle\dot{\vec{r}}\rangle$ (and so are their magnitudes). Namely, velocities align. 

It should be noted, additionally, that Eq.~\eqref{eq:mf} leads to describing the difference between individual velocities as
\begin{equation}
    (\dot{\vec{r}}_i - \dot{\vec{r}}_j) = (\vec{f}_i - \vec{f}_j) - \alpha_i\vec{u}_i + \alpha_j\vec{u}_j,
    \label{eq:spat}
\end{equation}
which neglects the covariance term introduced in Eq.~\eqref{eq:extavg}. Shifting our viewpoint, we can recognize that this expression also characterizes the evolution of the distance between individuals. In turn, it will enable us to obtain insights into the system's long-term spatial behavior.

\subsection{Coupling homogeneity}\label{sec:couphomo}
Although the main results of this paper rely on the system governed by Eq.~\eqref{eq:mf}, setting $\alpha_i = \alpha_j = \bar{\alpha}$ allows us to introduce the simpler form:
\begin{equation}
    \dot{\vec{r}}_i = \vec{f}_i + \bar{\alpha}\left(\langle\vec{r}\rangle - \vec{r}_i\right).
    \label{eq:simp_mf}
\end{equation}
Then, one can easily recognize that
\begin{equation}
    \langle\dot{\vec{r}}\rangle =\langle\vec{f}\rangle,
    \label{eq:simpavg}
\end{equation}
which does not depend on a covariance term, in contrast to Eq.~\eqref{eq:extavg}. This relation is particularly important, as it enables rewriting Eq.~\eqref{eq:simp_mf} as 
\begin{equation}
\dot{\vec{u}}_i = \vec{f}_i - \langle\vec{f}\rangle - \bar{\alpha}\vec{u}_i.  
\label{eq:flucdyn}
\end{equation}
Therefore, since $(\vec{f}_i - \langle\vec{f}\rangle)$ is constant, velocity fluctuations are coupled only through $\bar{\alpha}$. That is, if the coupling strength does not depend on the state of neighboring individuals, their velocity fluctuations are uncoupled. This peculiarity will be discussed in more detail towards the end of this paper; for now, it suffices to note that Eqs.~\eqref{eq:flucgen} and~\eqref{eq:flucdyn} are fundamentally different.

The homogeneity constraint, introduced in Eq.~\eqref{eq:simp_mf}, enables us to describe the system's behavior in a more intuitive fashion. Notice, from Eq.~\eqref{eq:flucdyn}, that choosing $\bar{\alpha}>0$ will always drive the system towards a \textit{bounded} state: $\vec{r}_i$ will be \textit{pulled} by $\langle\vec{r}\rangle$. Despite this \textit{attractive} effect, however, the only scenario in which $\vec{r}_i = \langle \vec{r} \rangle$ can occur is yielded by $\vec{f}_i = \vec{0}$, representing the collapse of all individuals into a static point.  In contrast, when $\vec{f_i}\neq\vec{0}$, the individual distance towards the center of mass ($\vec{u}_i$), converges to $\left[\vec{f}_i - \langle\vec{f}\rangle\right]/\bar{\alpha}$. That is, cohesion is stable and individuals display an \textit{apparent repulsion} from the center of mass. Remarkably, given this spatial convergence, the system's linear size (i.e., the maximum Euclidean distance between two individuals within the group) and its coupling strength, are expected to relate as $L_{\bar{\alpha}}\sim{\bar{\alpha}}^{-1}$.

Although one might expect similarities on the behavior of systems governed by Eqs.~\eqref{eq:mf} and~\eqref{eq:simp_mf}, the effects imprinted by the coupling heterogeneity cannot be neglected. Additional proof on this matter, also important for subsequent discussion, is found by analyzing the system's coherence. Thus, by using Eq.~\eqref{eq:simpavg}, we obtain:
\begin{equation*}
    |\langle\dot{\vec{r}}\rangle|^2 = |\langle\vec{f}\rangle|^2,
\end{equation*}
which suggests that, if coupling strengths are equal, the system's order relies exclusively on $\vec{f}$. That is, we can have full control over the system's coherence, unlike Eq.~\eqref{eq:op01} whose definition also depends on $\vec{r}$.

\subsection{Behavioral imperfections}
Although not explicitly described so far, behavioral rules are continuously at play during the system's evolution. Each individual, driven by a self-propelling velocity, seeks to follow its own trajectory, yet, this is disrupted by the \textit{attraction} exerted by the population's center of mass. During the transient, this dichotomy leads to the emergence of non-zero $\dot{\vec{u}}_i$, which then fade gradually as the population approaches the equilibrium. Steady cohesion manifests, in general, for a positive \textit{constant} coupling strength, as it guarantees that a \textit{balance} between behavioral rules will be achieved effectively.  Once settled, shifting the population's reference frame to that of its center of mass reveals that velocity fluctuations have vanished, as per the result of alignment shown earlier ($\dot{\vec{r}}_i = \langle\dot{\vec{r}}\rangle$). 

In pursuit of both convenience and the demonstration of the system’s robustness, we impose perturbations to individual responses; that is, we enforce \textit{behavioral imperfections}. To this end, we define the coupling strengths as
\begin{equation}
    \alpha_i = \alpha + \zeta_i(t),
    \label{eq:agg_noise}
\end{equation}
where $\alpha$ and $\zeta_i(t)$ are respectively constant and random variables. Notably, independent of the statistics underlying the latter, as long as $\alpha_i>0$, individuals will be \textit{pulled} by the center of mass, as glimpsed earlier (condition that holds even if $\alpha_i$ becomes temporarily negative). 

It is worth noting that there are notable differences between the noise we use (multiplicative) and \textit{standard} noise used in thermal systems (additive). A random variable added to Eq.~\eqref{eq:simp_mf} would represent perturbations that directly affect individual velocities. Moreover, if such noise is characterized by a zero mean, the system's average behavior would become exactly the same as in the case analyzed in Sec.~\ref{sec:couphomo}. In contrast, our definition [Eq.~\eqref{eq:agg_noise}] represents individual errors when following the rules which maintain the system cohesive. In addition, it is noteworthy that this multiplicative form maintains the system's heterogeneity ($\alpha_i \neq \alpha_j$) without significant complications.

\section{Collective states}\label{sec:sec03}
For the numerical results presented in this paper, we consider individuals moving in three dimensions. The random variable, from Eq.~\eqref{eq:agg_noise}, is characterized by an \textit{amplitude} $D_\alpha$ and statistics: $\langle\zeta_i(t)\rangle = 0$ and $\langle\zeta_i(t)\zeta_j(t')\rangle = D_\alpha^2\delta(t-t')\delta_{i, j}$. It must be noted that $D_\alpha$ and $D_\alpha^2$ are not the actual noise intensity; given its multiplicative nature, the \textit{effective intensity} depends on $\vec{u}_i$. To define the constant self-propelling velocities, individual components of $\vec{f}_i$ are drawn from uniform distributions between respective bounds $\vec{f}_{min}$ and $\vec{f}_{max}$ (see Table~\ref{tab:rang}). Additional computation details are left for Appendix~\ref{app:numerics}.

\begin{table}
\begin{tabular}{ccccccc}
\hline
        & \multicolumn{2}{c}{$x$-axis} & \multicolumn{2}{c}{$y$-axis} & \multicolumn{2}{c}{$z$-axis} \\ \hline
        & $f_{min}$        & $f_{max}$       & $f_{min}$        & $f_{max}$       & $f_{min}$        & $f_{max}$       \\ \hline
$f^{(1)}$ & 0            & 13          & 0            & 10          & -5           & 5           \\ \hline
$f^{(2)}$ & -10          & 10          & -10          & 10          & -5           & 5           \\ \hline
\end{tabular}
\caption{Bounds for uniform distributions used to draw individual self-propelling velocities ($\vec{f}_i$).}
\label{tab:rang}
\end{table}

Considering ($D_\alpha = 0$) leads the system to operate in the regime of homogeneous coupling strengths. Under this assumption, Eq.~\eqref{eq:conv} enables estimating the long-term distances between individuals as: $|\vec{r}_i - \vec{r}_j| = |\vec{f}_i - \vec{f}_j|/\alpha$, which, in turn, allows us to characterize the system's linear size as $L_\alpha \approx |\vec{f}_{\max} - \vec{f}_{\min}|\alpha^{-1}$. These expressions, however, do not hold when the coupling strengths are heterogeneous. Given that $\alpha_i \neq \alpha_j$, the effects of $\langle \vec{f}\rangle$ and $\mathrm{Cov}(\alpha, \vec{r})$ do not cancel out [Eq.~\eqref{eq:conv}]. In this scenario, therefore, the system's linear size ($L$) is expected to deviate from the baseline predicted by $L_{\alpha}$. This effect is displayed in Fig.~\ref{fig:fig01}, where $L$ is calculated in simulations for different values of $\alpha_i$.

\subsection{Spatial coherence}
\begin{figure}
    \centering
    \includegraphics[width=\linewidth]{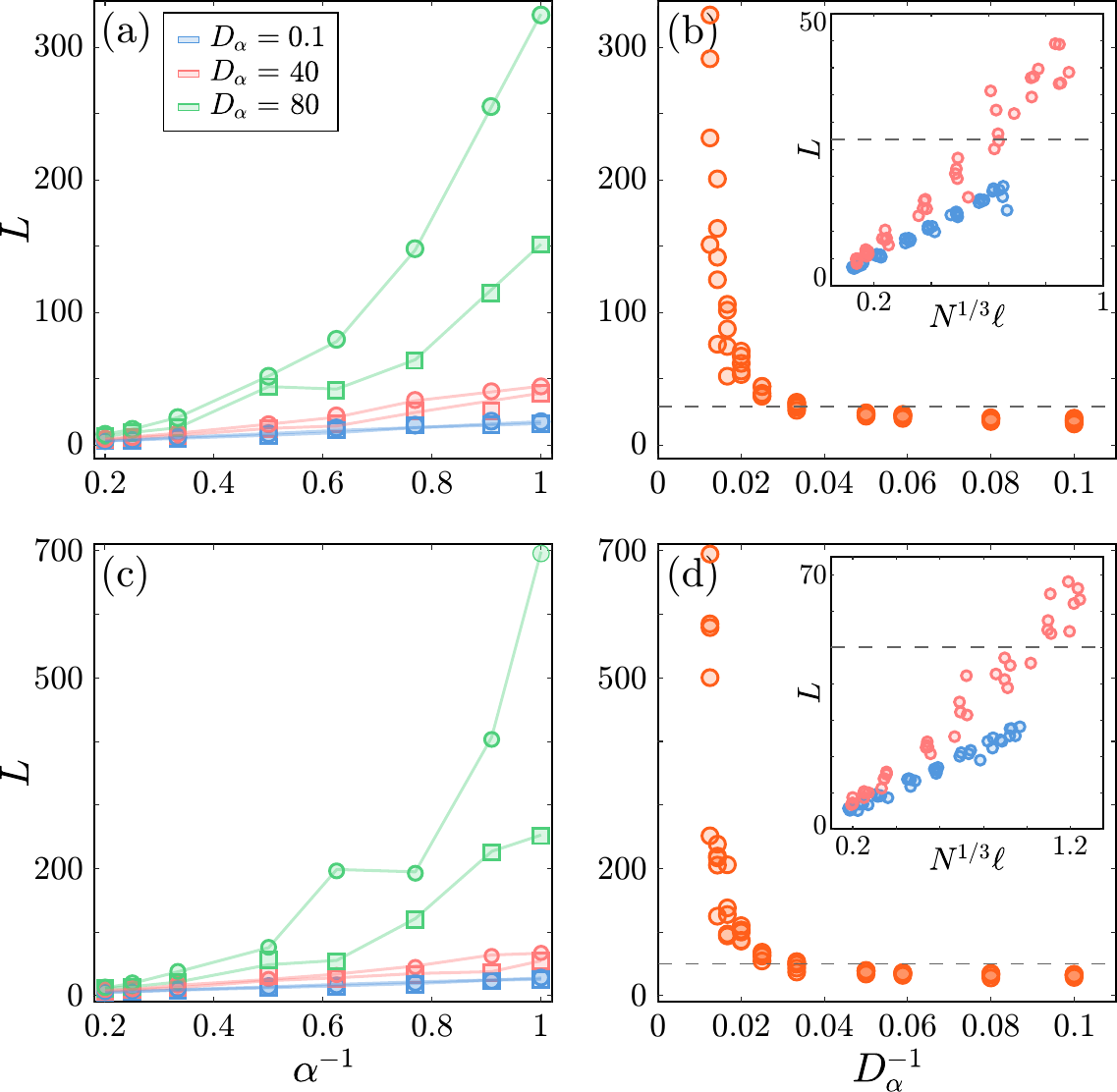}
    \caption{System's long-term linear size as a function of different coupling strengths (left column) and noise \textit{amplitudes} (right column). The self-propelling velocities were chosen according to Table~\ref{tab:rang}: $f^{(1)}$ (top row) and  $f^{(2)}$ (bottom row). Simulations corresponding to the left panels were performed for systems composed of $50$ (squares) and $500$ (circles) individuals. For the right panels, $\alpha$ was fixed at $1$; for their respective insets, however, the coupling strengths ($\alpha$) were chosen within $[1, 5]$. In both, figures and insets (right column), the number of individuals ($N$) is uniformly distributed between $50$ and $500$. The dashed lines are positioned at the same values of $L$ in each panel. See SM1 and SM2 in~\cite{SupMat} for movies (detailed descriptions are presented in Appendix~\ref{app:movs}).}
    \label{fig:fig01}
\end{figure}

From Figs.~\ref{fig:fig01}(a) and~(c), one may notice that as long as the noise \textit{amplitude} is small ($D_\alpha\rightarrow0$), the relation between $L$ and $\alpha^{-1}$ remains approximately linear. Remarkably, in this regime, the impact of the system's population size ($N$) on its linear size ($L$) is negligible. When the noise \textit{amplitude} experiences a moderate increment (see $D_\alpha = 40$), the linear relationship between $L$ and $\alpha^{-1}$ breaks, and, despite the emergence of a (quasi) linear regime, the effects of $N$ become noticeable. After surpassing an \textit{amplitude} threshold (see $D_\alpha = 80$), even the (quasi) linearity breaks, and $N$ starts to play a major role in the behavior of $L$.

Despite not explicitly displayed, Figs.~\ref{fig:fig01}(a) and~\ref{fig:fig01}(c) suggest that the system experiences a state transition. Furthermore, given that large variations in $L$ are triggered by the noise \textit{amplitude}, we infer that $D_\alpha$ acts as the system's control parameter. Figs.~\ref{fig:fig01}(b) and~\ref{fig:fig01}(d) allow us to focus on the noise's effects, as simulations were performed for a fixed $\alpha$. As displayed, up to a threshold ($D_\tau$), the system's linear size is not considerably affected by variations in $D_\alpha$; indeed, $L$ remains almost constant. Then, once the noise \textit{amplitude} surpasses the threshold ($D_\alpha > D_\tau$), even slightly, the linear size ($L$) grows abruptly.

Based on the observations described above, we can certainly confirm that the system experiences a state transition. Accordingly, since we found this phenomenon by evaluating the system's linear size, we deem it appropriate to characterize both ends of the transition in terms of the system's spatial distribution. In this vein, the insets of Figs.~\ref{fig:fig01}(b) and~\ref{fig:fig01}(d) show that, as long as $D_\alpha$ is \textit{small} ($D_\alpha<D_\tau$), the linear relationship $L\sim N^{1/3}\ell$ holds. Consequently, given the geometry of the system, we can infer that the mean inter-particle distance scales with the density as $\ell \sim \rho^{-1/3}$. That is, the particles are distributed (quasi) homogeneously in space. Once $D_\alpha$ drives the system along the state transition ($D_\alpha\geq D_\tau$), this linear relationship breaks, and so does the homogeneous spatial distribution. Notably, if we consider the limiting case where $D_\alpha\rightarrow0$, the spatial homogeneity and the scaling relation $L\sim\alpha^{-1}$, give rise to 
\begin{equation}
    \alpha\ell\sim N^{-1/3},
    \label{eq:al}
\end{equation}
which will be shown to be our system's defining feature.

\subsection{Orientational coherence}
In systems of individuals who possess orientations, the expected effect of large-\textit{amplitude} noise is that of disrupting their collective alignment. Therefore, as $D_\alpha$ was found to drive the system through a phase transition in space, it is reasonable to expect a close relation with a transition between polarization (alignment) states. 

\begin{figure}
    \centering
    \includegraphics[width=\linewidth]{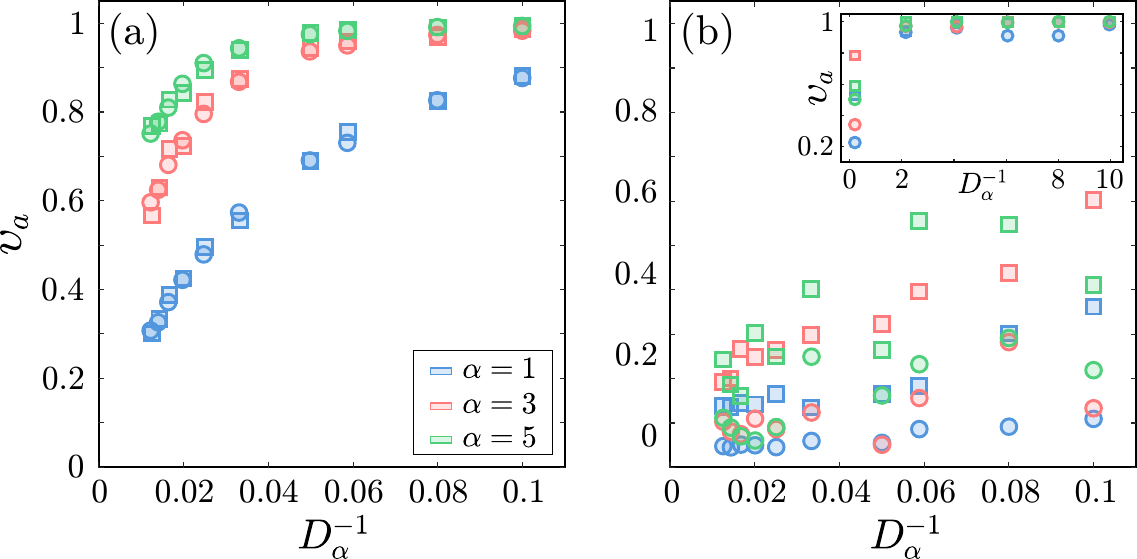}
    \caption{Average polarization of the system as a function of different coupling strengths and noise \textit{amplitudes}. The self-propelling velocities were drawn using $f^{(1)}$ for (a), and  $f^{(2)}$ for (b). All the simulations were performed for populations of $N = 50$ (squares) and $N = 500$ (circles) individuals. The inset complements (b) by extending its domain to include $D_\alpha$ values from $0.1$ to $5$; the abrupt decay occurs at $D_\alpha^{-1}\approx0.2$.}
    \label{fig:fig02}
\end{figure}

To measure the polarization of our system, we use the well-known order parameter $v_a = N^{-1}\left|\sum_i\left(\dot{\vec{r}}_i/|\dot{\vec{r}}_i|\right)\right|$~\cite{vicsek1995novel}, which describes a normalized version of Eq.~\eqref{eq:op01}. As expected, increasing the noise \textit{amplitude} ($D_\alpha$) results in the decrease of the polar order ($v_a$) [Fig.~\ref{fig:fig02}]. Notably, since $D_\alpha$ does not depend on $\alpha$, once the latter becomes larger, the effects generated by the noise become less significant (as long as the noise \textit{amplitude} is fixed). In spite of this, it is clear that the variations in $\upsilon_a$ depend primarily on $D_\alpha$, while the transition it experiences follows a shape partially outlined by $\langle\vec{f}\rangle$, in consistency with Eq.~\eqref{eq:op01}.

The most intriguing feature of the system is exposed when comparing Figs.~\ref{fig:fig01}(d) and ~\ref{fig:fig02}(b). According to the latter, when $D_\alpha\approx20$, the polarization is already low ($\upsilon_a<0.4$). Yet, in the former, the spatial phase transition does not occur until the noise \textit{amplitude} reaches $D_\alpha\approx 40$. That is, there is an intermediate regime where the system has already transitioned from collective alignment to non-alignment, while individuals remain distributed homogeneously.

\section{Effective criticality}\label{sec:sec04}
To delve deeper into the nature of the system's phase transition, we now focus on its expected behavior when $D_\alpha=D_\tau$. More precisely, on its macroscopic features after performing a Finite Size Scaling (FSS) analysis. Given that the tools we use in this section are \textit{standard}~\cite{amit2005field}, we omit their derivations for brevity.

\subsection{Scale invariance}
Our starting point is to define the correlation function
\begin{equation*}
    C(\epsilon) = \frac{\sum_{ij}\vec{\sigma}_i \cdot\vec{\sigma}_j \delta(\epsilon - r_{ij})}{\sum_{ij}\delta(\epsilon - r_{ij})},
\end{equation*}
where $r_{ij} = |\vec{r}_i - \vec{r}_j|$, and $\vec{\sigma}_i = \dot{\vec{u}}_i/\sqrt{\langle|\dot{\vec{u}}|^2\rangle}$. Since our system is finite, once it operates near-criticality, velocity fluctuations are expected to be positively correlated up to a threshold. Thus, we can define the zero-crossing correlation length as: $C(\epsilon = \xi_0) = 0$.

\begin{figure}
    \centering
    \includegraphics[width=\linewidth]{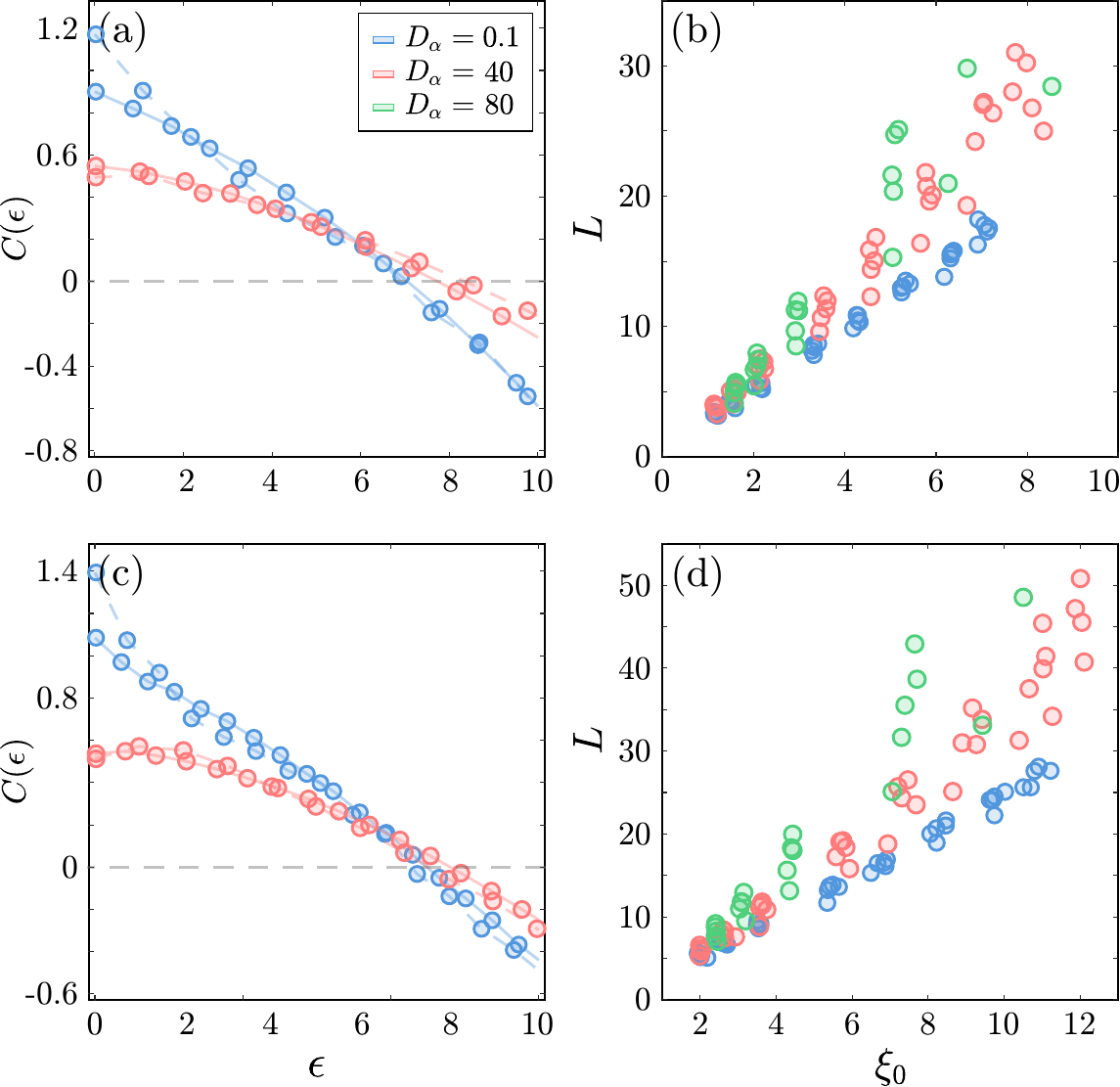}
    \caption{Scale invariance in the system for events with $\alpha= 1$ (left column), and $\alpha \in [1, 5]$ (right column). The self-propelling velocities were set following $\vec{f}^{(1)}$ (top row) and $\vec{f}^{(2)}$ (bottom row). For panels on the left, simulations were performed for $50$ (dashed lines) and $500$ (solid lines) individuals, and the $\epsilon$-axis is cropped. For those on the right, the population sizes were chosen as: $N\in[50,500]$. For the sake of clarity, the cases for $D_\alpha = 80$ are shown only in panels on the right. In these, moreover, the axes fully cover only the points corresponding to cases with $D_\alpha = 0.1$ and $D_\alpha = 40$. For larger sizes ($L$), the trend of the curve corresponding to $D_\alpha=80$ is to bend.} 
    \label{fig:fig03}
\end{figure}

Figs.~\ref{fig:fig03}(a) and~\ref{fig:fig03}(c) show the correlation functions computed for the same systems evaluated in Figs.~\ref{fig:fig01} and ~\ref{fig:fig02}. As displayed, independent of the self-propelling velocities or the number of individuals, velocity fluctuations are found to be positively correlated up to a spatial threshold ($\xi_0$). Notably, the most significant difference between correlation curves is that zero-crossing points ($C(\epsilon) = 0$), for considerably different noise \textit{amplitudes} ($\Delta D_\alpha\approx 40$), show an offset that is barely perceptible. This observation is particularly important, as the system's polarization is expected to decrease as $D_\alpha$ increases [Fig.~\ref{fig:fig02}]. That is, Fig.~\ref{fig:fig03}(a) and~\ref{fig:fig03}(c) demonstrate that positive velocity correlations are present independent of the polarization exhibited by the system.

Complementing the correlation curves, Figs.~\ref{fig:fig03}(b) and~\ref{fig:fig03}(d) illustrate the dependence of the system's correlation length ($\xi_0$) on its linear size ($L$). Remarkably, for cases where $D_\alpha\leq40$, the system exhibits scale invariance. Namely, as the system's size increases, the size of the domains where individual velocities correlate positively scales as: $\xi_0\sim L$. 

\subsection{Polarization crossover}
In the finite-size regime, the emergence of scale-free correlations ($\xi_0\sim L$) is \textit{typically} linked to systems operating in a near-critical state. However, when collective coherence relies on the alignment of individual velocities, near-criticality features can also be exhibited far from the critical point. Once individuals point in the same orientation, the system's rotational symmetry breaks; in turn, transversal correlations appear indicating the existence of Goldstone modes. 
Given that our system exhibits scale invariance in states of \textit{high} and \textit{low} polarization, one could conjecture that both origins might coexist (near-criticality and Goldstone modes), as in other models dealing with the same phenomenon. Still, to make an accurate conclusion, we evaluate the system's susceptibility, $\chi$.

In the thermodynamic limit, $\chi$ diverges when the system operates in criticality. However, in finite-size systems, for each linear size ($L$) (or $N$ given the spatial homogeneity), a specific pseudocritical point ($\tau_N$) marks the value of the control parameter in which the susceptibility reaches a finite maximum. Following the FSS hypothesis, thus, the evolution of these susceptibility peaks is characterized by

\begin{equation}
    \chi\sim(\tau_N - \tau_c)^{-\gamma},
    \label{eq:susc0}
\end{equation}
where $\gamma$ is a critical exponent, and $\tau_c$ is the critical point. 

\begin{figure}
    \centering
    \includegraphics[width=\linewidth]{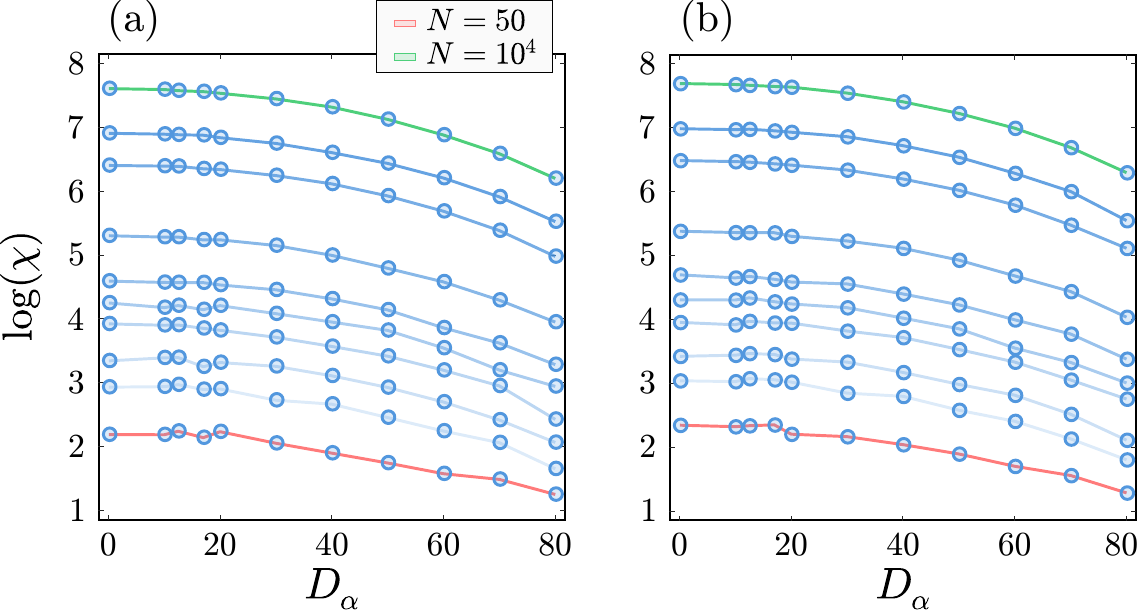}
\caption{Static crossover in the transition between the system's states of \textit{high} and \textit{low} polarization. The coupling strength $\alpha$ is fixed at $1$, and the population sizes range from $N = 50$ to $N = 10^4$. The increment of $N$ is represented by the blue curves becoming bolder.}
    \label{fig:fig04}
\end{figure}

To measure the susceptibility in our system, we use the proxy~\cite{attanasi2014finite}
\begin{equation*}
    \chi = \frac{1}{N}\sum_{ij}\vec{\sigma}_i\cdot\vec{\sigma}_jH(\xi_0 - r_{ij}),
\end{equation*}
where $H(.)$ is the Heaviside function. Strikingly, by considering the noise \textit{amplitude} as the system's control parameter ($\tau=D_\alpha$), the relationship described in Eq.~\eqref{eq:susc0} is not fulfilled [Fig.~\ref{fig:fig04}]. No traces of pseudocritical points (susceptibility peaks) are observed, even when the population size is increased. In contrast, as $D_\alpha$ changes gradually, the susceptibility is found to evolve smoothly, in what appears to be a static crossover between states of \textit{high} and \textit{low} polarization. 

As displayed before, scale invariance is also present in states of \textit{low} polarization. Therefore, despite the apparent absence of a critical point when $\tau = D_\alpha$, we cannot state with certainty that the system's scaling feature emerges due to the presence of Goldstone modes. Building on this, we may venture to suggest the existence of another control parameter which does enable the presence of pseudocritical points.

\subsection{Structural criticality}
Within the FSS framework, the susceptibility of a system operating near-criticality can be defined as 
\begin{equation*}
    \chi \sim L^{\gamma/\nu},
\end{equation*}
where $\nu$ is a critical exponent. Then, by equating this expression to Eq.~\eqref{eq:susc0}, we can obtain
\begin{equation}
    (\tau_N - \tau_c)\sim L^{-1/\nu},
    \label{eq:susc1}
\end{equation}
which describes the size-dependence of the distance between pseudocritical and critical points. Evidently, based on previous observations [Fig.~\ref{fig:fig04}], we can infer that taking $\tau_N = D_\alpha$ would not satisfy Eq.~\eqref{eq:susc1}. Our goal, then, is to identify the parameter that does. To this end, we focus on the limit where $D_\alpha\rightarrow0$, which, furthermore, implies that the system exhibits spatial homogeneity [Fig.~\ref{fig:fig01}]. 

\begin{figure}
    \centering
    \includegraphics[width=\linewidth]{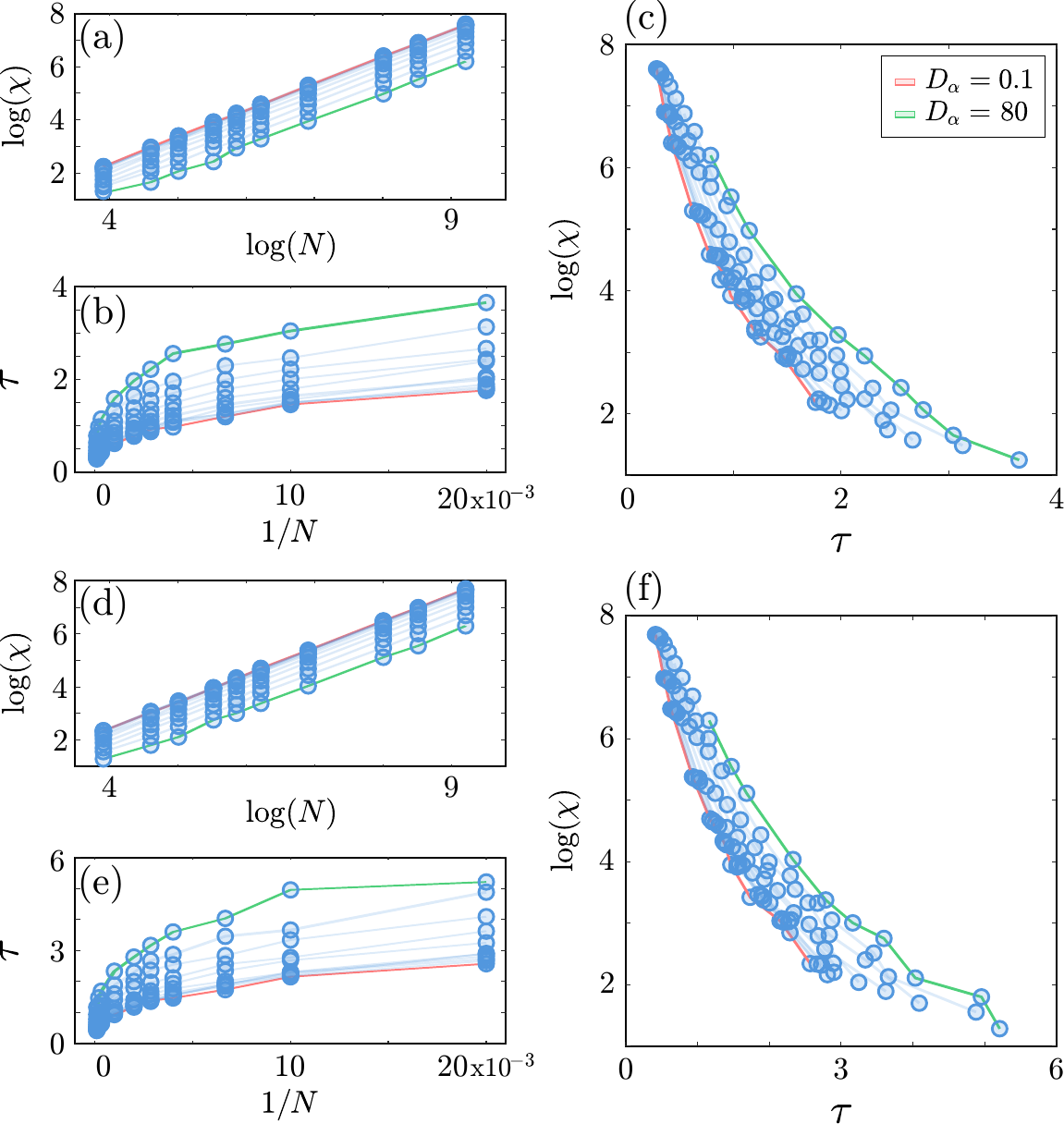}
    \caption{Susceptibility of the system considering the mean-interparticle distance as the control parameter ($\tau = \ell$). All simulations were performed for $\alpha = 1$ and population sizes between $N=50$ and $N = 10^4$. The self-propelling velocities were chosen using ${f}^{(1)}$ for the top panels [(a), (b), and (c)], and ${f}^{(2)}$ for the bottom ones [(d), (e), and (f)]. Blue lines characterize the susceptibility calculated for intermediate values of $D_\alpha$, between $0.1$ and $80$.}
    \label{fig:fig05}
\end{figure}

Strictly speaking, since the left-hand side of Eq.~\eqref{eq:susc1} depends on $N$, the right-hand side must also show this dependence; yet, this is not explicitly defined. To address this, we start by considering the system subject to a constant $\alpha$. In this scenario, according to Eq.~\eqref{eq:al}, the only varying parameter is $\ell$ (and consequently $\rho$). Then, in the thermodynamic limit ($N\rightarrow\infty$), we expect to have $\ell_{N\rightarrow\infty} = 0$. Once we explicitly define the spatial homogeneity relation:
\begin{equation*}
\ell\sim N^{-1/3}L,
\end{equation*}
a direct mapping to Eq.~\eqref{eq:susc0} becomes evident. The control parameter is identified as the mean interparticle distance ($\tau = \ell$), implying that $\tau_c = \ell_{N\rightarrow\infty}$. Since Eq.~\eqref{eq:al} dictates that $\ell \sim L$ for a fixed $N$, the system is \textit{structurally constrained} to satisfy the scaling relation $(\tau - \tau_c) \sim L^{-1}$, consistent with Eq.~\eqref{eq:susc1} for $\nu=-1$ (if $\tau = \rho$, then $\nu = 1/3$). That is, the \textit{structural restriction} [Eq.~\eqref{eq:al}] establishes that $\ell$ as a control parameter \textit{self-regulates} to satisfy the scaling relations that characterize the system's near-critical state [Fig.~\ref{fig:fig05}]. 

While Fig.~\ref{fig:fig05} shows that scaling relations are consistent for $\tau = \ell$, this control parameter provides a size-specific description. As exhibited in Fig.~\ref{fig:fig05}, the effects of noise become noticeable only when its \textit{amplitude} surpasses a threshold. More precisely, the offsets arise from variations of the system's size (which evolves according to Fig.~\ref{fig:fig01}). Similarly, offsets are expected to arise when comparing events characterized by different $\alpha$ (even for $D_\alpha = 0$). Therefore, to achieve a \textit{universal scaling}, we need to define a normalized control parameter: $\tau \sim \alpha\ell$ (divided by $|\langle\vec{f}\rangle|$ to make it dimensionless). Thus, the scaling curves corresponding to different $\alpha$ collapse.

It is worth remarking that, while mathematically Eq.~\eqref{eq:al} represents the homogeneity condition ($L\sim N^{1/3}\ell$), it physically acts as an \textit{equation of state}. Unlike \textit{standard} systems, where $L$ is fixed as a boundary condition independent of the system's dynamics, in our case $L$ is dynamically determined by the coupling strength ($L\sim\alpha^{-1}$). That is, Eq.~\eqref{eq:al} implies that the system's density (through $\ell$) is \textit{locked} to the coupling parameter $\alpha$, thus linking the system's geometry with its governing dynamics.

\section{Discussion}\label{sec:sec05}
We have presented a minimal model of aggregation in which individual velocities align as a byproduct of global cohesion. Given the model's simplicity, we derived expressions regarding the system's stability and long-term order. Notably, the system is characterized by scale invariance, which, through an FSS analysis, we found to be a \textit{structural} rather than an emergent feature. Finally, a striking regime was found to emerge, where the system's orientational and spatial coherence are decoupled. More precisely, even when the collective exhibits \textit{low} polarization, the distribution of positions remains homogeneous ($\ell\sim\rho^{-1/3}$).

The coupling heterogeneity ($\alpha_i \neq \alpha_j$) was shown to play an important role in the system's behavior, as it couples the fluctuation dynamics [Eq.~\eqref{eq:flucgen}]. To preserve its effects, we found it convenient to split $\alpha_i$ so that noise is \textit{internal} rather than \textit{external} (additive). Apart from mathematical convenience, this definition [Eq.~\eqref{eq:agg_noise}] was inspired by nature, where achieving a \textit{perfect balance} between behavioral rules is not the norm. Individual errors in obeying these, even while nearly aligned, give rise to non-zero $\dot{\vec{u}}_i$~\cite{cavagna2010scale}, also referred to as \textit{inherent noise}~\cite{yates2009inherent}. 

In models within the \textit{traditional} velocity-alignment framework, the system's collective state is usually controlled by fixing its density and the \textit{amplitude} of a perturbing noise. Accordingly, beyond models, natural swarms have also been shown to exhibit emergent collective states triggered by the same two parameters~\cite{attanasi2014finite,attanasi2014collective}. In our system, given the effects generated by the noise \textit{amplitude} ($D_\alpha$) on the polarization, we found it natural to identify it as a control parameter. However, by evaluating the system's susceptibility, we found that the transition driven by $D_\alpha$ does not correspond to a \textit{traditional} second order transition, but rather to a crossover between states of \textit{high} and \textit{low} polarization. On the other hand, we demonstrated that, as long as spatial homogeneity holds, the density (as a control parameter) fulfills the scaling relations characteristic of pseudocritical points. It is worth recalling that we do not set the system's density, but it evolves to maintain consistency with the governing dynamics. A similar phenomenon has been observed in the self-organization of midges, where the system's density \textit{self-tunes} into criticality~\cite{attanasi2014finite}.

Our study has been limited to constant self-propelling velocities, which impose a sort of \textit{quenched} disorder into the system. Put simply, using the form introduced in Eq.~\eqref{eq:simp_mf}, the system's equilibrium results from a balance between individual ``stubbornness'' ($\vec{f}_i$) and ``social conformity'' ($\bar{\alpha}$). Then, in a loose sense, the collective ends up behaving as a \textit{flying crystal}. Once coupling strengths are considered heterogeneous, the system's equilibrium exhibits its dependence on a covariance term [Eq.~\eqref{eq:conv}]. This subtle feedback describes an emergent self-regulation mechanism: if conformists (high $\alpha_i$) are positioned on the group periphery, the system's mean velocity will be reduced; in contrast, if conformists gather near its center of mass, it will accelerate. A related phenomenon was observed in fish schools~\cite{puy2024selective}, where individual speeds determine the emergence of \textit{temporal leaders}. Remarkably, the heterogeneity characterizing our system (whether through $\vec{f}_i$ or $\alpha_i$) aligns with the relevance of individual differences for the \textit{optimal} functioning of animal and artificial collectives~\cite{jolles2017consistent, jolles2020role, teng2022heterogeneity, montanari2025optimal}.

As presented in Eq.~\eqref{eq:op01}, the system's orientational coherence depends partially on the average self-propelling velocity. Based on this, it is trivial to recognize that, if $\langle\vec{f}\rangle  \approx 0$, the polar coherence depends exclusively on the covariance term: $|\langle\dot{\vec{r}}\rangle|^2 \approx  \mathrm{Cov}(\alpha, \vec{r})^2$. This effect has been evidenced in Fig.~\ref{fig:fig02}(b), as $f^{(2)}$ corresponds to distributions with zero mean in each axis (see Table~\ref{tab:rang}). For a noteworthy remark, we now shift our focus to Eq.~\eqref{eq:spat}, which characterizes the system (if not unstable) actively trying to suppress the evolution of interparticle distances: $(\dot{\vec{r}}_i-\dot{\vec{r}}_j) \rightarrow 0$. Notably, this tendency is maintained even when $\langle\vec{f}\rangle\approx 0$, conflicting with the expected disorder ($|\langle\dot{\vec{r}}\rangle|^2\rightarrow0$) under the same consideration. These contrasting effects thus outline the decoupled nature of the system's spatial and orientational coherence.

A rich scenario arises when self-propelling velocities are time-dependent. As previously noted from Eq.~\eqref{eq:mf}, the convergence of the system towards its equilibrium depends on $\alpha_i$ being positive. This, however, guarantees a stable behavior in the simplest case: when individual trajectories are linear. An improvement to that condition would be that of $\alpha_i$ being strong enough for individuals to follow specific non-linear paths. To illustrate this, let us assume that self-propelling velocities are defined in such way that collective motion is spiral. This behavior is, indeed, more similar to natural swarms than the states obtained using $f^{(2)}$ (see SP~\cite{SupMat}). Then, if the relaxation time is too slow (small $\alpha_i$), individuals will find themselves trying to follow the reference spiral trajectory, but failing. As a result, velocity fluctuations will emerge without the need for a source of noise (whether additive or multiplicative). Indeed, this phenomenon also constitutes a type of \textit{behavioral imperfection}, generating effects analogous to that of multiplicative noise [Eq.~\eqref{eq:agg_noise}]. 

The scenario in which self-propelling velocities are time-dependent is especially significant, as it allows for the definition of the governing dynamics: $\dot{\vec{f}}_i (t)$. Notably, despite the complexity of $\dot{\vec{f}}_i(t)$, the system's instantaneous average behavior is expected to be described by the same equations derived in this paper. Consider, for instance, that we do not rely on multiplicative noise, but fluctuations emerge solely by effects of $\vec{f}_i(t)$ (as described for spiral paths). Consequently, coupling strengths become homogeneous ($\alpha_i = \alpha_j$), and, at any given instant, fluctuations will uncouple [Eq.~\eqref{eq:flucdyn}]. This setting leaves $\vec{f}_i(t)$ as the only term on which locality can be imposed, effectively distinguishing between driving (local) and cohesion (global) mechanisms.

It should be noted that the notion of \textit{locality} in our model differs fundamentally from that in alignment-based models. As a general example, in \textit{classical} frameworks describing magnetic systems, individuals are usually distributed on nodes of discrete lattices. Spatial self-organization becomes secondary, and thus local interactions are solely responsible for aligning neighboring \textit{spins}. Even when individuals possess the capacity for spatial movement, locally aligned clusters recombine by means of additional considerations such as periodic boundary conditions~\cite{vicsek1995novel, supekar2023learning}. Thus, the system's spatial distribution ends up relegated to a secondary role. In our aggregation-based system, in contrast, collective coherence depends primarily on how individuals self-organize in space. Given that particles move freely in open space, locality would drive the emergence of small clusters that would not recombine, as each would adopt a preferred orientation. In this sense, staying close to the entire collective is more important than just staying close to their immediate neighbors. 

Although the aggregation term formally couples all individuals through the global center of mass, the resulting dynamics naturally confine the population inside a finite spatial domain. Within this cohesive group, interactions effectively act over a limited range, as distant individuals contribute almost uniformly to the mean field. In this sense, locality is not imposed by construction (as in alignment-based models) but emerges as a consequence of spatial cohesion. 

While our model's mean-field nature is not strictly realistic, its value lies in its simplicity. This minimalist approach, allows us to capture foundational physics, namely the \textit{structural criticality} and \textit{coherence-decoupling}, that more complex models fail to reproduce. More properly, we consider our framework to be a baseline for system-specific complexity. An example of this claim is presented in~\cite{lizarraga2025collective}, where we introduce an underdamped extension of Eq.~\eqref{eq:mf} which reproduces specific information-transfer features exhibited by natural flocks.

Although further investigation is required, it is worth mentioning that numerical simulations were also performed for $\vec{f}_i$ sampled from normal distributions, with the system's fundamental behavior remaining unaffected. As a final remark, even though the results presented in this paper focus on individuals moving in three dimensions, the model introduced in Eq.~\eqref{eq:mf} describes a system without such restriction. 
\begin{acknowledgments}

\end{acknowledgments}

\bibliography{apssamp}
\appendix
\section{Numerics}\label{app:numerics}
All simulations were performed over $5000$ time-steps, and long-term averages were calculated using the last $4000$ samples (corresponding to the state after the transient). The numerical integration of Eq.~\eqref{eq:mf} was computed using Heun's method with time-steps of $\Delta t = 0.01$. For its implementation, we define 
\begin{align*}
    \begin{split}
        p(\vec{\kappa}_i) &= \vec{f}_i + \alpha(\langle\vec{\kappa}\rangle - \vec{\kappa}_i),\\
        q(\vec{\kappa}_i) &= D_\alpha(\langle\vec{\kappa}\rangle - \vec{\kappa}_i),
    \end{split}
\end{align*} for any $\vec{\kappa}_i$. Then, the predictor obeys
\begin{equation*}
    \vec{r}_i^{(*)} = \vec{r}_i^{(t)} + p(\vec{r}_i^{(t)})\Delta t + q(\vec{r}_i^{(t)})\eta_i^{(t)}\sqrt{\Delta t},
\end{equation*}
where $\eta_i^{(t)}\sim\mathcal{N}(0, 1)$, and the corrector:
\begin{equation*}
    \vec{r}_i^{(t+\Delta t)} = \vec{r}_i^{(t)} + \frac{1}{2}[p(\vec{r}_i^{(t)}) + p(\vec{r}_i^{(*)})]\Delta t + q(\vec{r}_i^{(t)})\eta_i^{(t)}\sqrt{\Delta t}.
\end{equation*}
No self-propelling velocities ($\vec{f}_i$) were fixed or reused; their values were randomly calculated at each execution of the simulations. 

The system's correlation functions and susceptibilities were calculated using expressions presented in the analysis of real systems~\cite{cavagna2010scale,attanasi2014collective, attanasi2014finite}. Our results, moreover, are consistent with those obtained using the second order definition (for correlations) and the variance of the order parameter (for susceptibility).

\subsection{Movies}\label{app:movs}
All movies correspond to simulations evaluated within the main text. Individuals are displayed distributed in a three-dimensional space. Blue and gray arrows represent, respectively, individual velocities and velocity fluctuations, both normalized. The view adapts---zooming in or out---to focus on the individuals; boundary conditions are not fixed. 

The prefixes SM1 and SM2 correspond to cases defined by $f^{(1)}$ and $f^{(2)}$, respectively (see Table.~\ref{tab:rang}). The prefix SP denotes the cases involving spiral trajectories, which we simulated inspired by the motion of real midges~\cite{cavagna2023characterization}. For this purpose, we defined the self-propelling velocities as 
\begin{equation*}
  \vec{f}_i(t) = (A\sin\omega_i t, A\cos\omega_i t, B\omega_{i,z}), 
\end{equation*}
where $A$ and $B$ (set to 1) are defined solely to ensure dimensional consistency. Similar to the cases where self-propelling velocities are constant, the frequencies defining $\vec{f}_i(t)$ in this scenario are drawn from uniform distributions: $\omega_i\in[5, 10]$ and $\omega_{i,z}\in[-5, 5]$. Additional parameters used for each movie are described in Table.~\ref{tab:params}.

\begin{table}[!ht]
\begin{tabular}{cccccc}
\hline
$D_\alpha$ & $0$ & $0.1$ & $10$                                               & $40$ & $80$ \\ \hline
SM1        &     & SM11  & SM12                                               & SM13 & SM14 \\ \hline
SM2        &     & SM21  & SM22                                               & SM23 & SM24 \\ \hline
SP         & SP1 &       & \begin{tabular}[c]{@{}c@{}}SP2\\ SP2x\end{tabular} & SP3  &      \\ \hline
\end{tabular}
\caption{For all movies, $\alpha$ and $N$ were fixed at $1$ and $50$, respectively. Although the simulations lasted for 5000 time steps, the movies display only the first 625 samples. In SP2x, particles are represented by circles rather than arrows.}
\label{tab:params}
\end{table}

\end{document}